\documentclass[11pt]{article} \usepackage{mystyle-new}
\usepackage{epsfig,amsmath} \usepackage{graphics}
\usepackage{cite,./mcite}
\usepackage{epsfig,amsmath} 
\usepackage{hepnames,hepunits}
\usepackage{hyperref}
\usepackage{subfigure}
\usepackage{float}
\hypersetup{
  colorlinks=true, 
    linktoc=all,     
    linkcolor=blue,  
    citecolor=red
}

                \def\lsim{\mathrel{\rlap{\lower4pt\hbox{\hskip1pt$\sim$}}
    \raise1pt\hbox{$<$}}}                \def\gsim{\mathrel{\rlap{\lower4pt\hbox{\hskip1pt$\sim$}}
    \raise1pt\hbox{$>$}}}

\newcommand{\pt}{\ensuremath{p_{\rm T}}}

\def\cascade{{\sc Cascade3}}

\def\pythia{{\sc Pythia}}

\usepackage{lineno}
\makeindex
\begin{document}
\begin{center} {\sffamily\Large\bfseries 
TMDs from Monte Carlo event generators  \\ \vspace*{0.2cm}
}
{ 
 \Large 
H.~Jung$^{1}$, S.~Steel$^{2}$,
S.~Taheri~Monfared$^{1}$, Y.~ Zhou$^{3}$ 

}  \vspace*{0.2cm}
     {\large $^1$DESY, Hamburg, Germany}\\
     {\large $^2$University of North Georgia, United States}\\
     {\large $^3$University of Cambridge, United Kingdom}

\end{center}

\begin{abstract}
Transverse Momentum Dependent (TMD) parton distributions are a very powerful concept for the description of low and high transverse momentum effects in high energy collisions.
The Parton Branching (PB) method provides TMD distributions which can be used in parton shower simulations, as already implemented in \cascade\ Monte Carlo event generator.

This report gives a description of the work done during the DESY summer student program 2021, young scientists from very different time zones connecting from remote twice a day for 8 weeks, to develop a method, PS2TMD, that allows to determine effective TMDs from the standard Monte Carlo parton showers. 
This method is validated and implemented to successfully reconstruct the PB-TMDs with different configuration settings.
We also discuss kinematic shifts in longitudinal momentum distributions from initial state showering and point out the sizable influence of different reconstruction definitions on both collinear and transverse momentum PDFs. 

\end{abstract}

\section{Introduction}
The cross section of high energy collisions is traditionally calculated as a convolution of a hard scattering process (calculable in perturbative Quantum Chromo Dynamics) with parton densities. Both the hard process and the parton densities are usually calculated in collinear factorization, meaning that only the longitudinal momenta of the colliding partons are considered, while the transverse momenta are entirely neglected. However, for a number of processes, in particular the \pt\ distribution of the Z boson at high energies, the prediction of purely collinear calculations leads to wrong results and a resummation of soft gluons to all orders has to be performed. This resummation can be performed analytically~\cite{Bizon:2018foh,Bizon:2019zgf,Catani:2015vma,Scimemi:2017etj,Bacchetta:2019tcu,Bacchetta:2018lna,Ladinsky:1993zn,Balazs:1997xd,Landry:2002ix,resbosweb,Alioli:2015toa,Bozzi:2019vnl,Baranov:2014ewa}, via the Parton Branching (PB) method~\cite{Martinez:2019mwt,Hautmann:2017xtx,Hautmann:2017fcj} or also by the simulation of parton showers in Monte Carlo event generators~\cite{Sjostrand:2014zea,Bellm:2015jjp,Bahr:2008pv,Gleisberg:2008ta} . \\
\\
Here, the PS2TMD approach developed in \cite{Schmitz:427383,Schmitz:2019krw} has been further extended to provide a set of Transverse Momentum Dependent (TMD) parton distributions which were entirely obtained from a simulation of parton showers of the Monte Carlo generator \pythia{8} . 

\section{Parton Branching method 
}

 Evolution equations such as DGLAP \cite{Lipatov:1975, Altarelli:1977,Gribov:1972ri, Dokshitzer:1977} , CCFM \cite{Ciafaloni:1988,Catani:1990,Catani:s1990,Marchesini:1995} , BFKL 
\cite{Kuraev:1976,Kuraev:1977,Balitsky:1978} describe the evolution of the parton densities with respect to different resolution scales.  The essential evolution equations pertaining to this Monte Carlo parton shower study is DGLAP. This PB  method provides an iterative solution to this evolution equation to retain and obtain kinematics while conserving energy and momentum for every branching. Moreover, PB describes the forward evolution towards the hard process from the smallest  to the largest energy scales ($\mu^2$), while Parton Showering (PS) is the equivalent reverse process from large energy scales to small. The PB method can be extended beyond just the collinear case, and can be applied to calculate TMD parton distributions as-well.  The evolution equations of the TMD parton density function, ${A}_{a}\left(x, \mathbf{k}, \mu^{2}\right)$ can be solved using the PB method.
 By applying the PB formalism to the evolution equation, the iterative solution for every resolvable branching is expressed as:
\begin{eqnarray}
\mathcal{A}_{a}\left(x, \mathbf{k}, \mu^{2}\right) &=\Delta_{a}\left(\mu^{2}\right) \mathcal{A}_{a}\left(x, \mathbf{k}, \mu_{0}^{2}\right)+\sum_{b} \int \frac{d^{2} \mathbf{q}^{\prime}}{\pi \mathbf{q}^{\prime 2}} \frac{\Delta_{a}\left(\mu^{2}\right)}{\Delta_{a}\left(\mathbf{q}^{\prime 2}\right)} \Theta\left(\mu^{2}-\mathbf{q}^{\prime 2}\right) \Theta\left(\mathbf{q}^{\prime 2}-\mu_{0}^{2}\right)  \nonumber \\
& \times \int_{x}^{z_{M}} \frac{d z}{z} P_{a b}^{(R)}\left(\alpha_{\mathrm{s}}, z\right) \mathcal{A}_{b}\left(\frac{x}{z}, \mathbf{k}+(1-z) \mathbf{q}^{\prime}, \mathbf{q}^{\prime 2}\right) + . . .~.
\end{eqnarray}

where a, b describe the particular parton flavor, $\Delta_{a}(\mu^2)$ is the Sudakov form factor which accounts for the probability of no emission or resolvable branching from the energy scale $\mu_{0}$ to $\mu$, and $P_{a b}^{(R)}$ is the DGLAP splitting kernel for real emissions \cite{Martinez:2019mwt }. In Reference \cite{Martinez:2019mwt}, precision measurements for inclusive HERA Deep Inelastic Scattering (DIS) were  fitted to the theoretical calculations using PB method. The resulting TMD PDFs are labelled PB-NLO-HERAI+II-2018-set1 (PB-set1) and PB-NLO-HERAI+II-2018-set2 (PB-set2). These two sets differ only by the choice of renormalization scales for the argument in strong coupling. This is taken to be either the evolution scale (in PB-set1) or the transverse momentum (in PB-set2).
The two PB-TMD sets can be integrated over $k_t$ to obtain their collinear PDFs (iTMDs), $f_i(x,\mu^2)$. For convenience, those collinear sets are called PB-set1 and PB-set2. They are required to generate hard processes.

\section{Effective TMDs from parton showerers }
Parton showering simulations with Monte Carlo generators provide a method, PS2TMD, to reconstruct TMD parton distributions. 
In this method, the cumulative effects of soft parton emissions to all orders are considered. 
Different \pythia{8} parton shower configurations are investigated and the resulting TMD parton distributions are studied.

\subsection{Validation of the method using PB-TMDs}
\begin{figure}[!h]
    \centering
    \includegraphics[width=6cm]{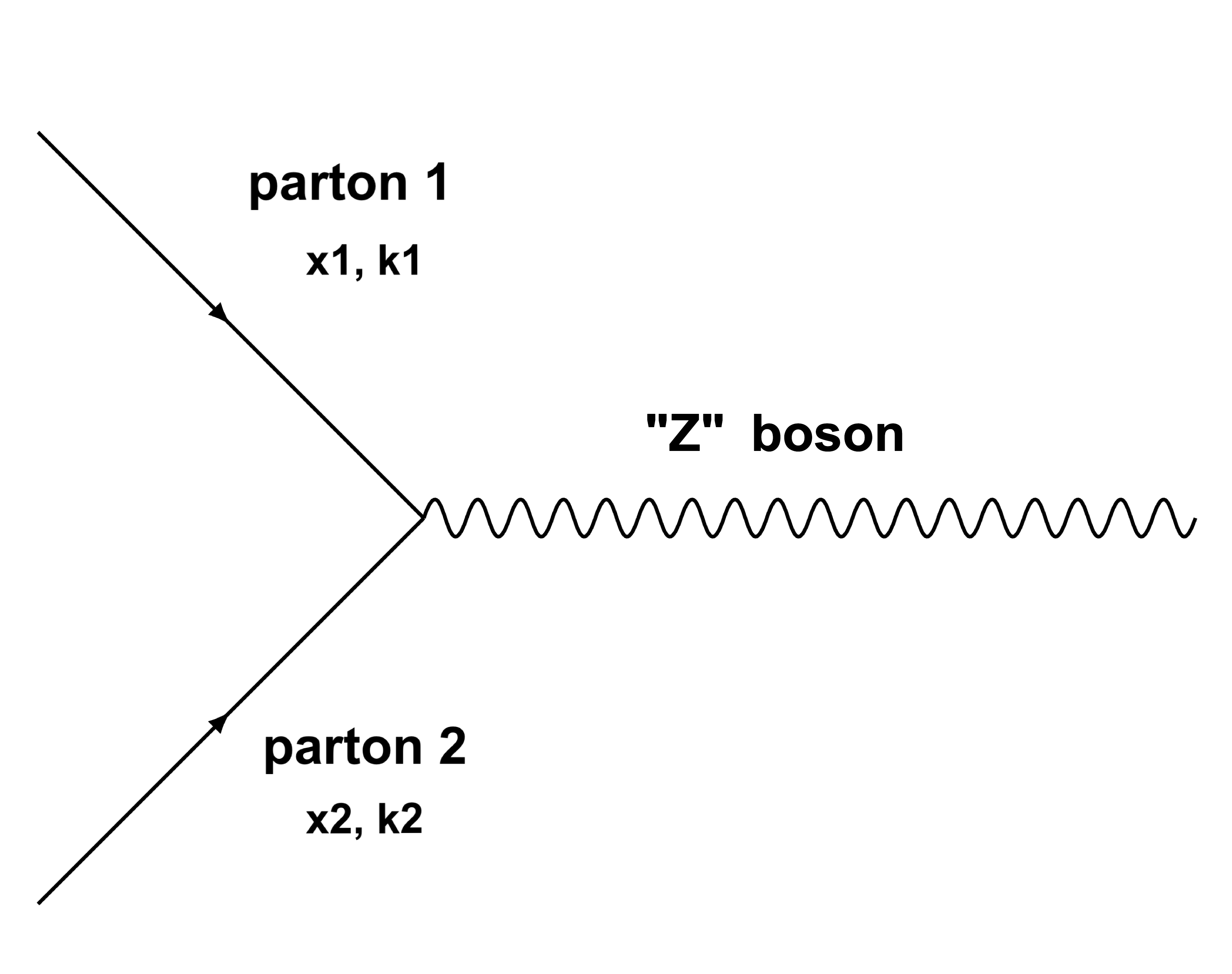}
    \caption{Feynman diagram of PS2TMD toy model.}
    \label{fig:PS2TMD-1}
\end{figure}

At the very beginning, the method, dubbed ``PS2TMD'', needs to be developed that enables the extraction of effective TMDs from parton shower. These TMDs can be calculated from a physical process. While due to kinematic limits, the determination of the quantities would be very complicated. We define a simple toy process for pseudo Z boson production, as shown in Figure \ref{fig:PS2TMD-1}, with 
the cross section only proportional to the parton density of parton 2. 
%
The initial state parton $k_1$ has a fixed momentum fraction close to one
(technically $x_1 = 0.99$ is used) and no transverse momentum for simplicity. 
Momentum conservation trivially gives $k_{2}=k_{Z}$.  Knowing \(k_{Z}\) from Monte Carlo generator event records, \(k_{2}\) 
can be calculated and histogrammed to obtain its TMD distribution. 
The hard process is simulated with \pythia{8} and the events are stored in the LHE format which can be read by parton shower event generators.
This is essentially how the PS2TMD method can be applied to  reconstruct TMDs from parton showers.

\begin{figure}[!h]
    \centering
    \subfigure[]{\includegraphics[width=0.42\textwidth]{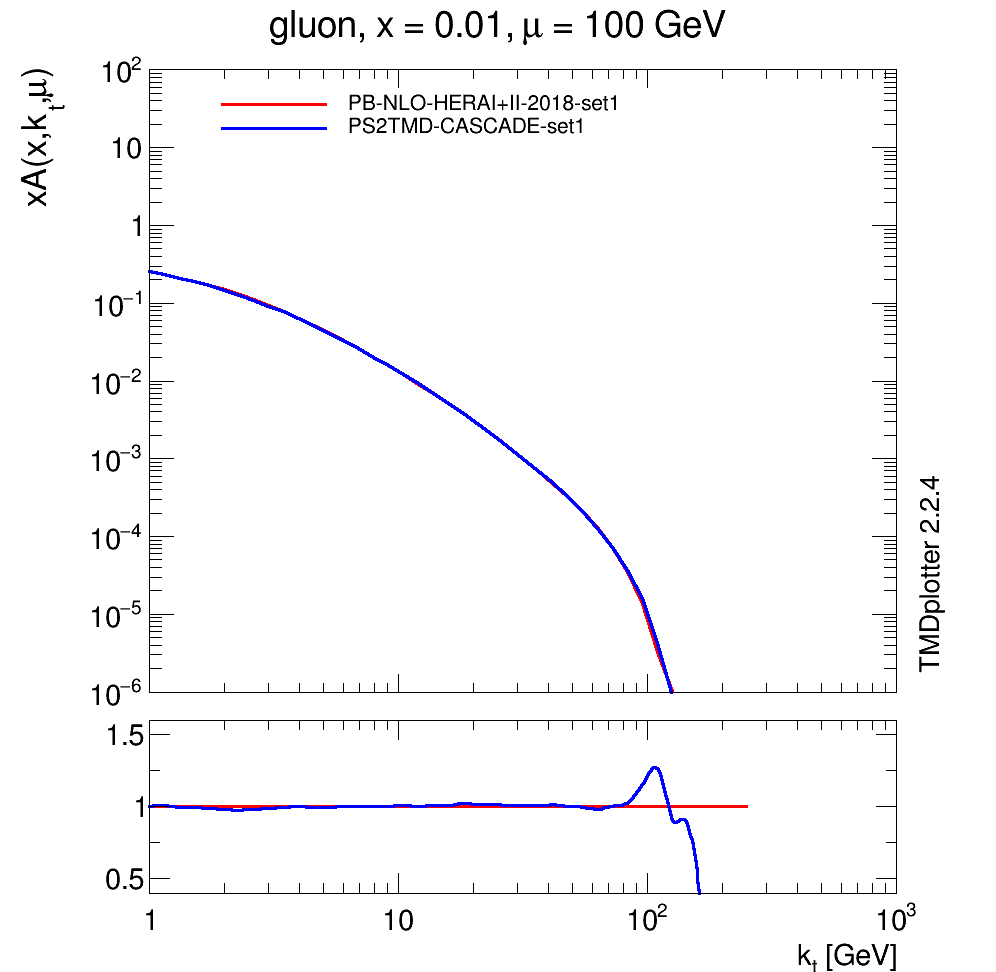}} 
    \subfigure[]{\includegraphics[width=0.42\textwidth]{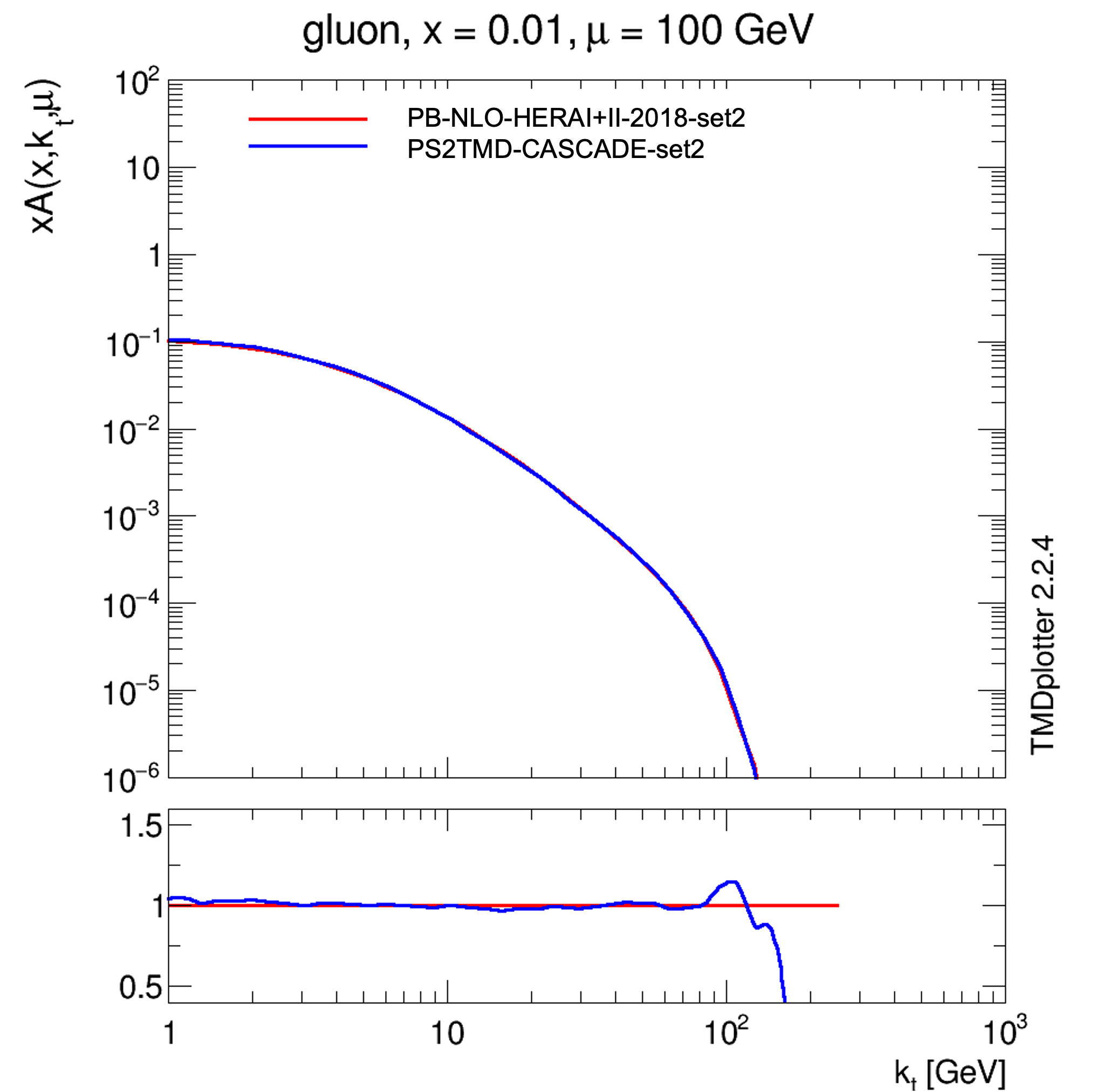}} 
    \subfigure[]{\includegraphics[width=0.42\textwidth]{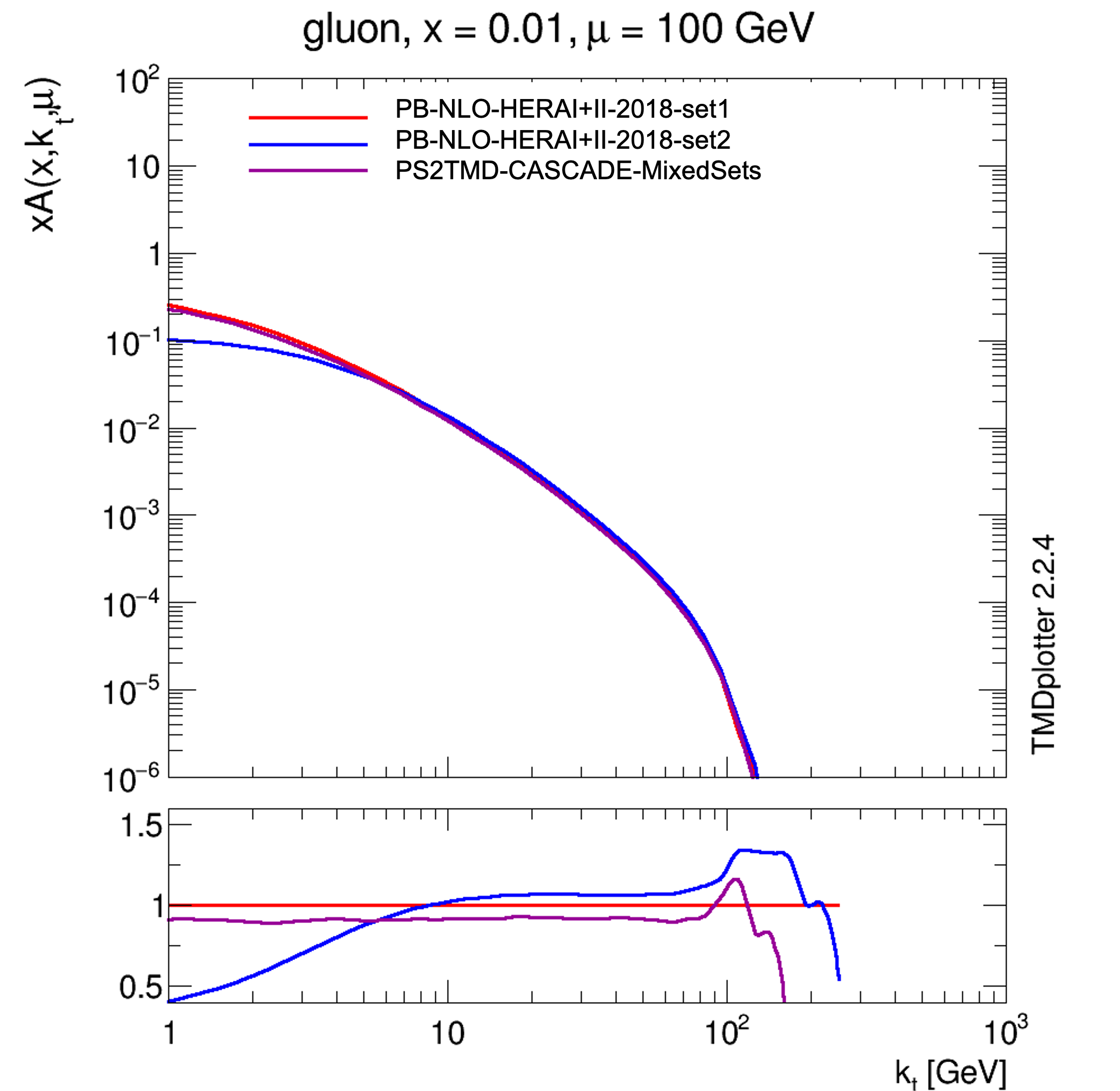}}
    \caption{(a) PS2TMD validation result for PB-set1. (b) PS2TMD validation result for PB-set2. (c) PS2TMD result when PB-Set1 and PB-TMD-Set2 are used.}
    \label{fig:valid-result}
\end{figure}

Before the main study, the PS2TMD method must be validated. 
To perform this cross check, TMDs are used instead of a parton shower. We apply the \cascade{} Monte Carlo generator and compare the results to the input distributions.
In Figure \ref{fig:valid-result} (a), collinear PB-set1 is used to generate hard processes and PB-TMD-set1 also fed into \cascade{}. This is done likewise in Figure \ref{fig:valid-result} (b) with collinear and TMD set2. It is evident from those two figures that when TMD sets are used consistently, the reconstructed TMDs agree with the initial TMDs within less than 2\% difference, validating PS2TMD method. It shows that indeed one can obtain a TMD distribution by analysing the final state of a MC event. Some small deviations at large $k_t$ are coming from statistical fluctuations.
In Figure \ref{fig:valid-result} (c), PB-set1 is used for hard processes while PB-TMD-set2 is fed to \cascade. We observe that the reconstructed TMDs disagree with both of the PB-sets.
Thus an inconsistent use of the PB-Sets for the collinear and TMD calculation does not reproduce the PB-Set that is applied for including the TMD.

The reconstructed TMDs are plotted with \textit{online TMD plotter} \cite{Abdulov:2021ivr}. Effective TMDs of \(k_t\) for gluons are given to best showcase the difference caused by parton shower configurations - similar patterns are observed in other parton flavors. 

\subsection{Effective TMDs from \pythia{8}}
In the following, effective TMDs are determined from the \pythia{8} parton shower (not from TMD as performed in the method validation).
\begin{figure}[!h]
    \centering
    \includegraphics[width=8cm]{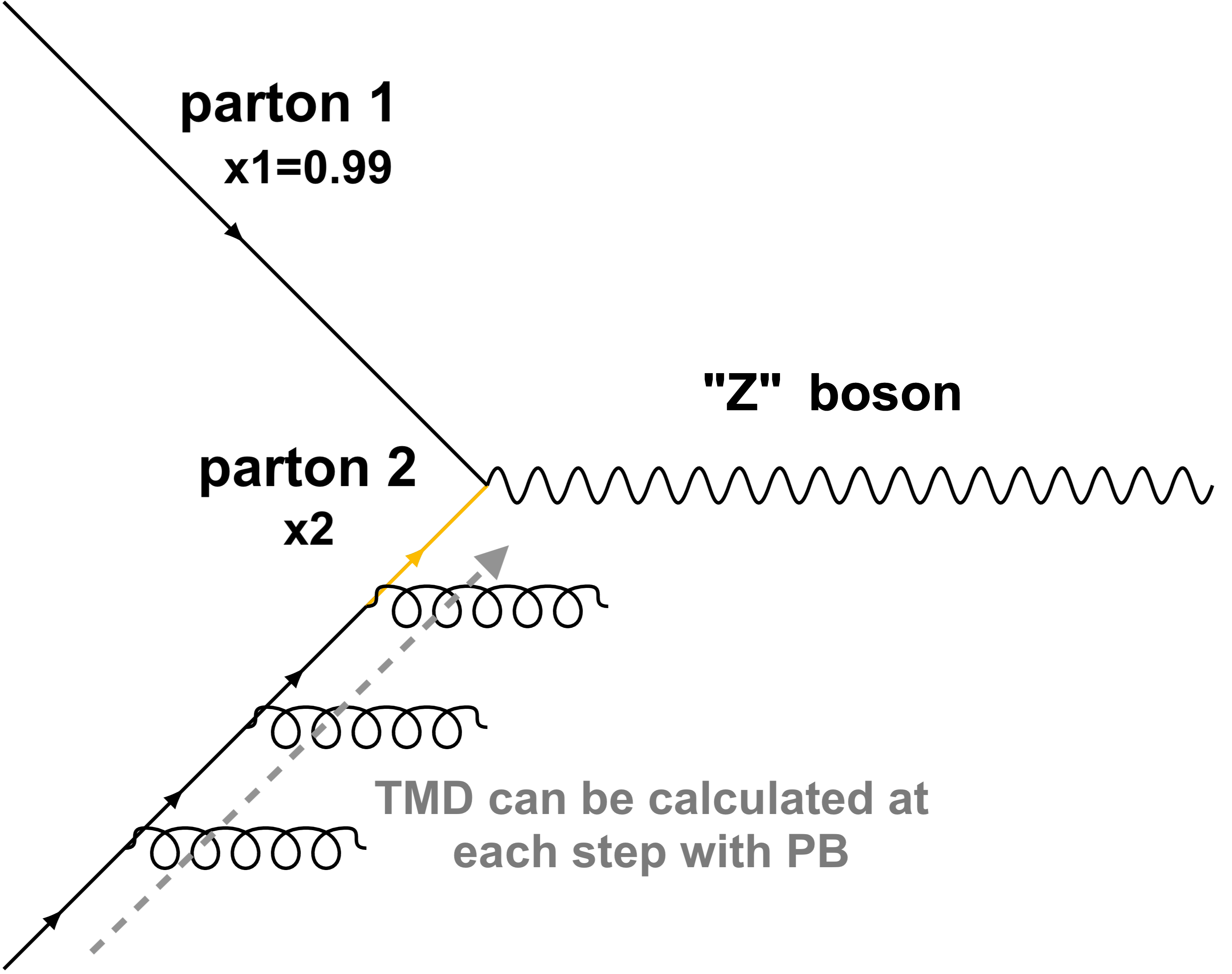}
    \caption{Feynman diagram of PS2TMD toy model when parton shower is included. The yellow parton is the one of which \(k_T\) distribution is studied. Curly lines attached to parton 2 represent parton emissions.}
    \label{fig:PS2TMD-2}
\end{figure}
As shown in Figure \ref{fig:PS2TMD-2}, parton shower is only considered on parton 2 in the form of initial state radiation (ISR). The energy of parton 1 is fixed to be 99\% of proton energy. The outgoing particle is set as Z boson for an easy identification in the event record. 
With Monte Carlo generators, one first simulates hard processes with the collinear PDFs, then generates a parton shower from initial and final state partons. Finally, one applies PS2TMD method to obtain the effective TMDs.
With the PS2TMD method validated, both steps can be invoked with \pythia{8} to study various configurations. 

Several switches with the most direct relevance to the parton shower were selected.



\begin{figure}[!h]
    \centering
    \subfigure[]{\includegraphics[width=0.42\textwidth]{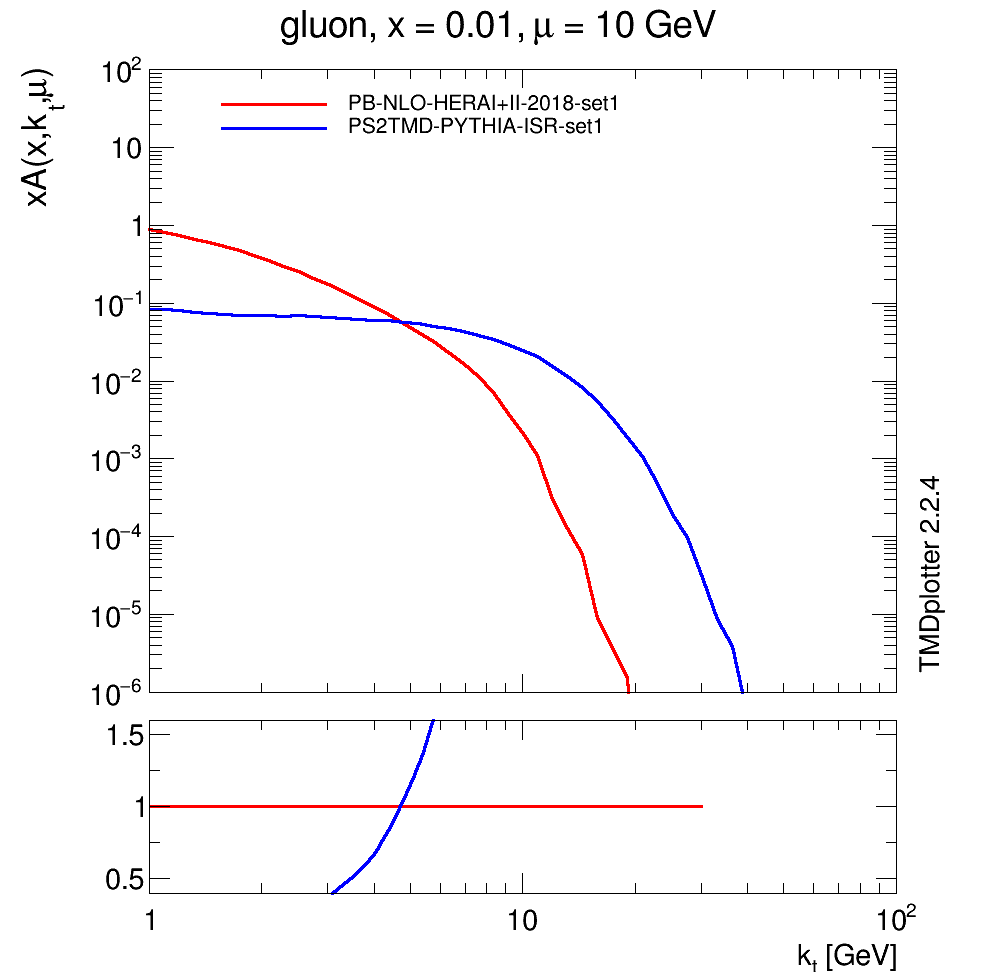}} 
    \subfigure[]{\includegraphics[width=0.42\textwidth]{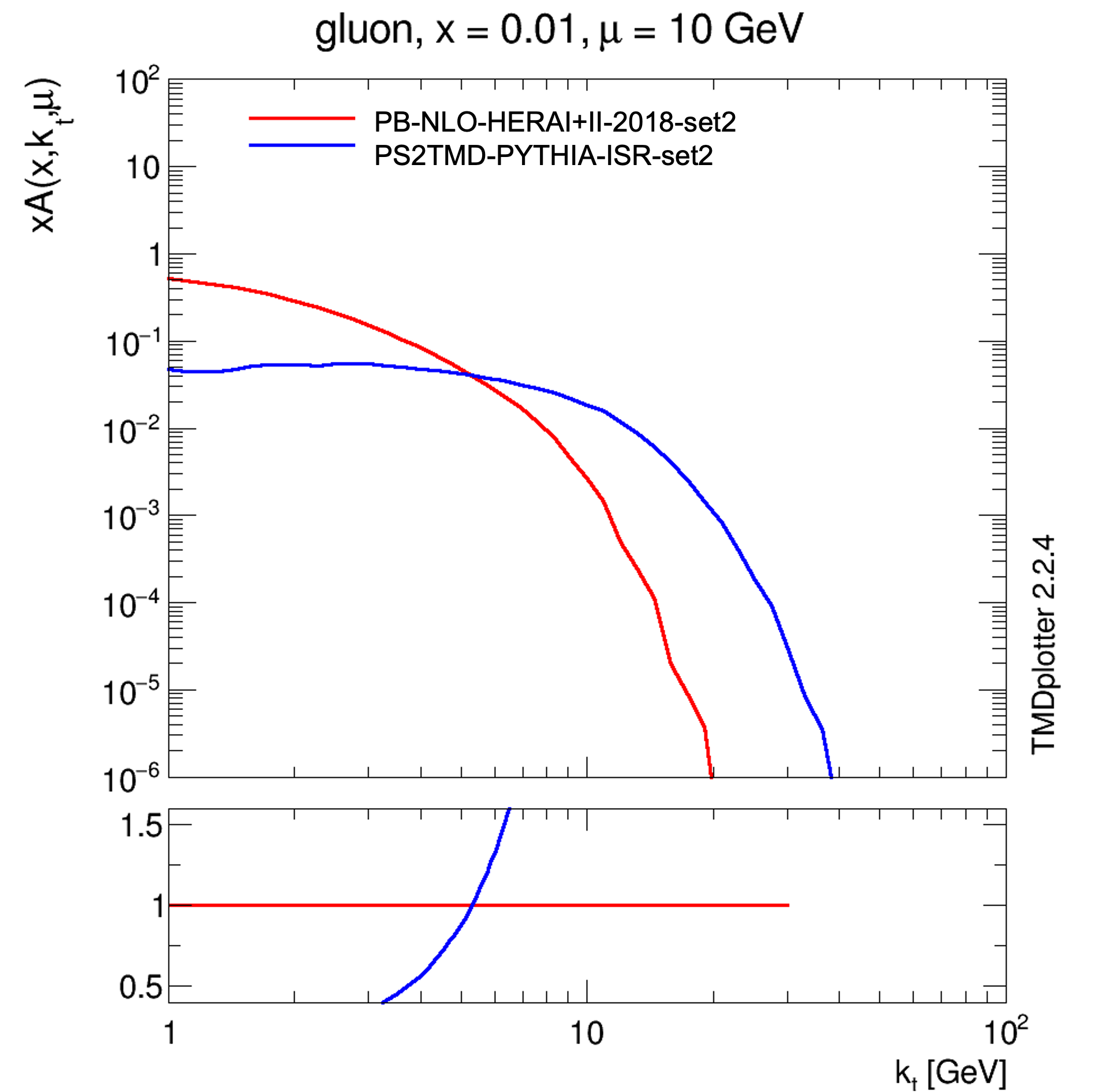}}
    \caption{Reconstructed \(k_T\) distribution from default parton showering for (a) PB-set1 and (b) PB-set2. Red line is PB-set1 and blue line is the reconstructed TMD from \pythia{8}.}
    \label{fig:ISR}
\end{figure}

\paragraph{ISR result} Effective TMDs from default parton shower configuration are obtained from both sets, shown in Figure \ref{fig:ISR}. The difference in shape between PB-TMDs and \pythia{8} reconstructed TMDs is attributed to the different ordering conditions they adopted. PB method uses angular ordering while \pythia{8} is based on \(p_T\) ordering. 

\paragraph{ISR + MPI and ISR + Primordial \(k_T\) OFF} We also investigated the effect of including Multi-Parton Interaction (MPI) and primordial \(k_T\) on \pythia{8} configuration. They result in a negligible effect on obtained TMDs. Following their definition, this is expected since they do not largely influence the hard processes. 

\begin{figure}[!h]
    \centering
    \subfigure[]{\includegraphics[width=0.42\textwidth]{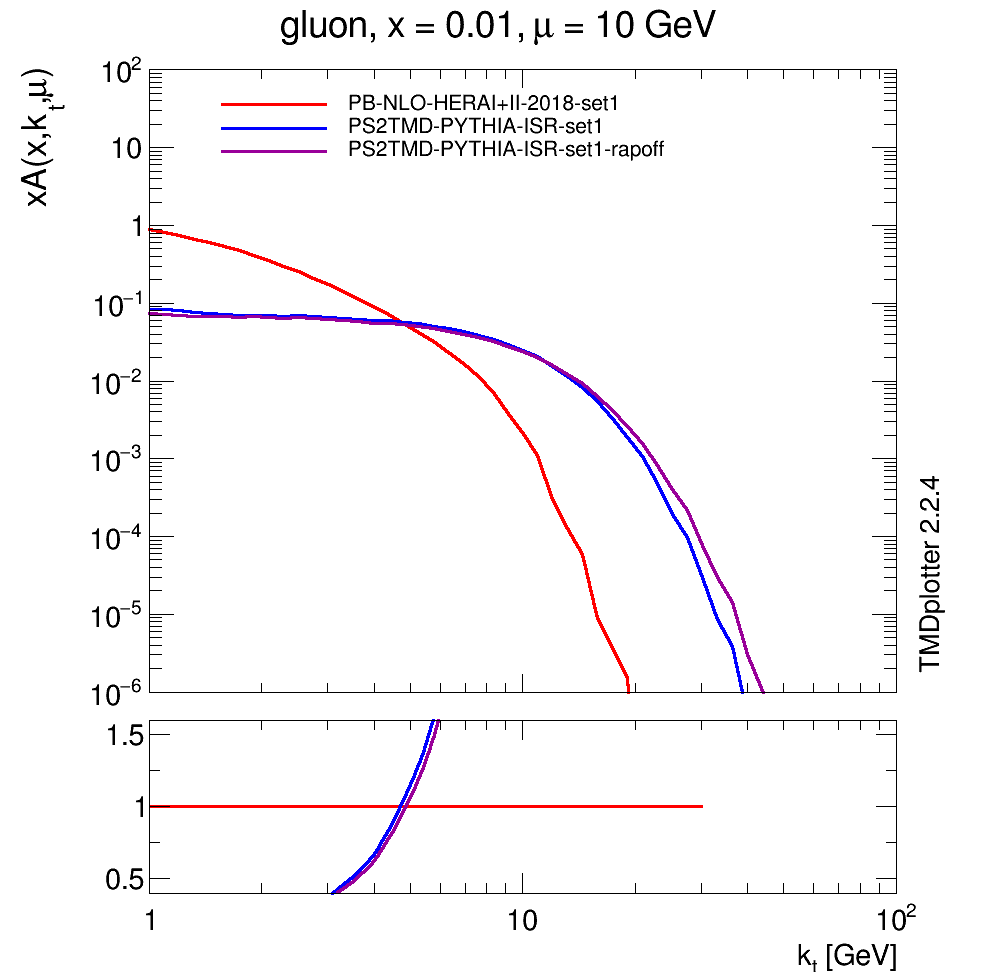}} 
    \subfigure[]{\includegraphics[width=0.42\textwidth]{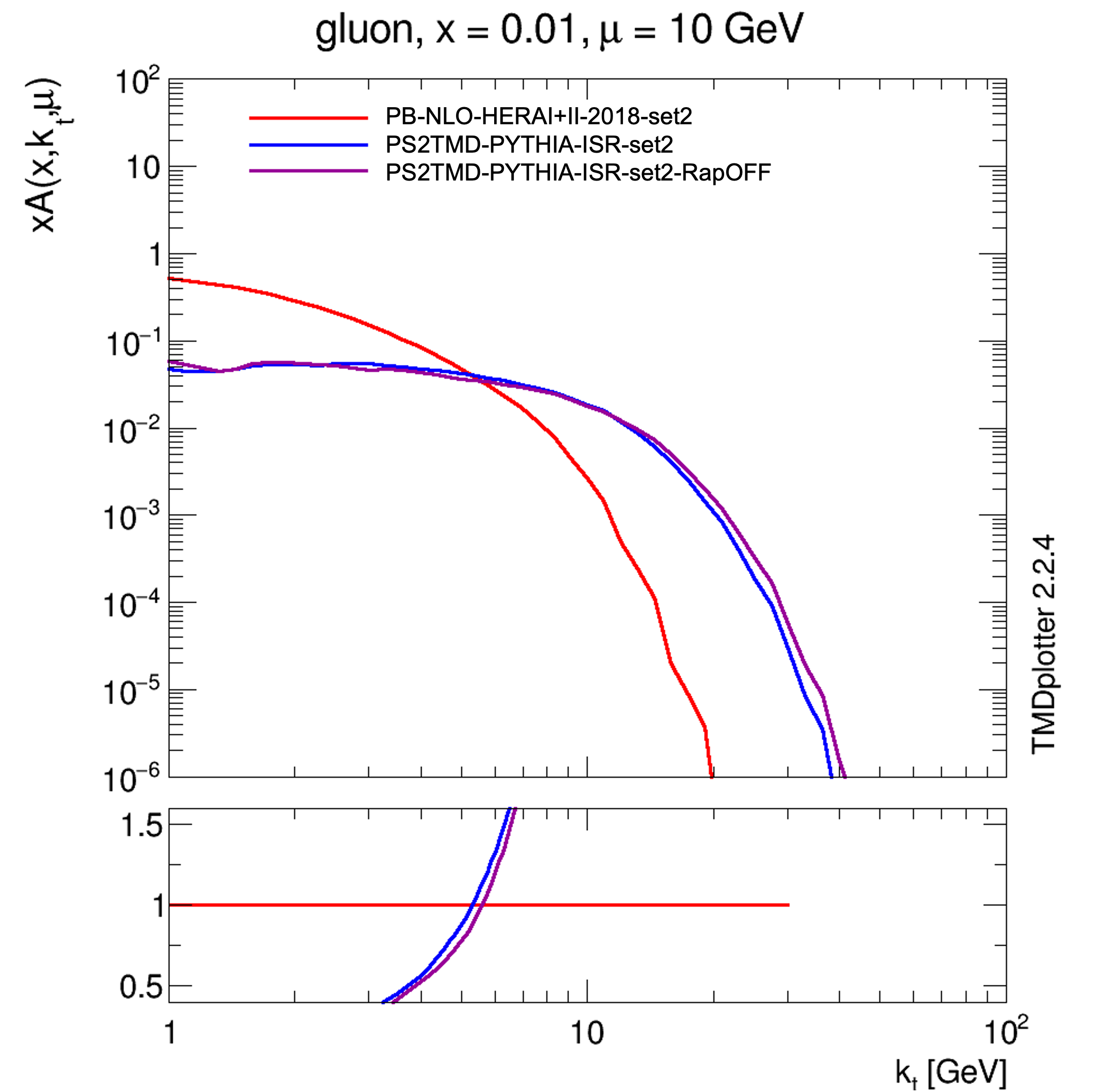}}
    \caption{Reconstructed \(k_T\) distribution when rapidity ordering is switched OFF for (a) PB-set1 and (b) PB-set2.}
    \label{fig:rap}
\end{figure}
\begin{figure}[!h]
    \centering
    \subfigure[]{\includegraphics[width=0.42\textwidth]{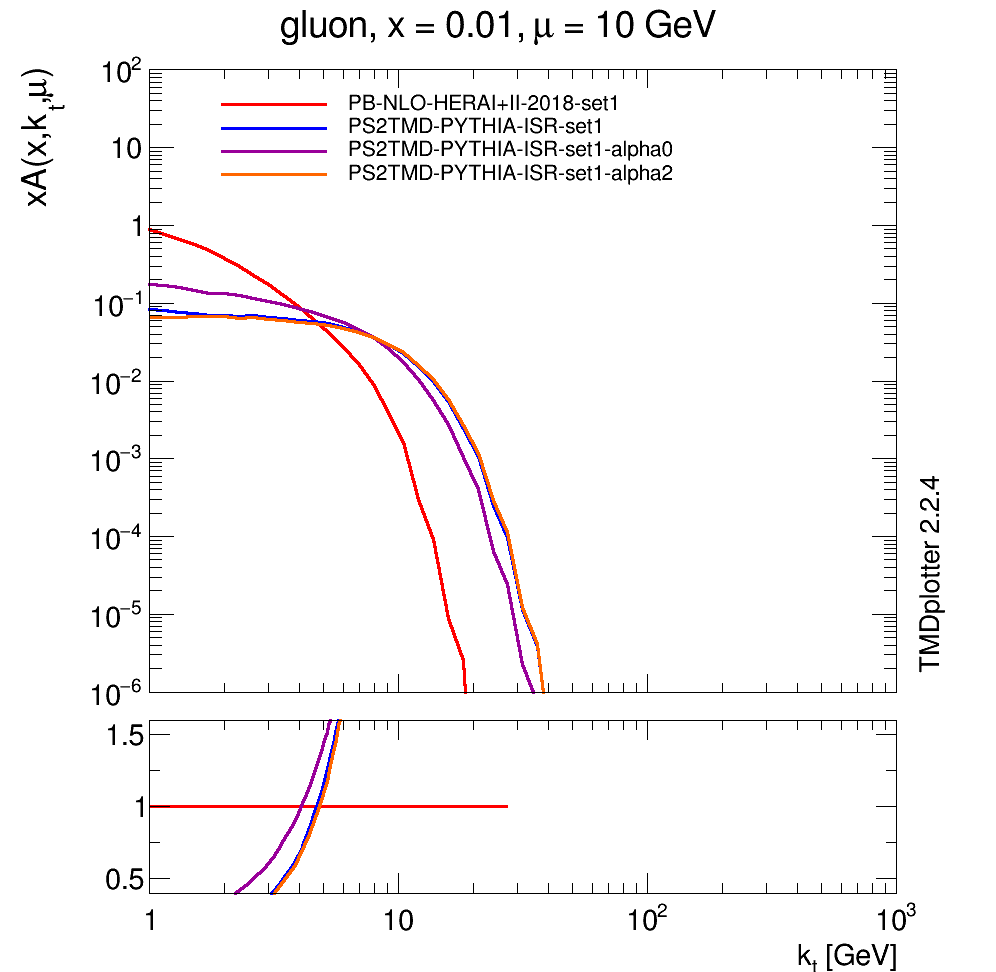}} 
    \subfigure[]{\includegraphics[width=0.42\textwidth]{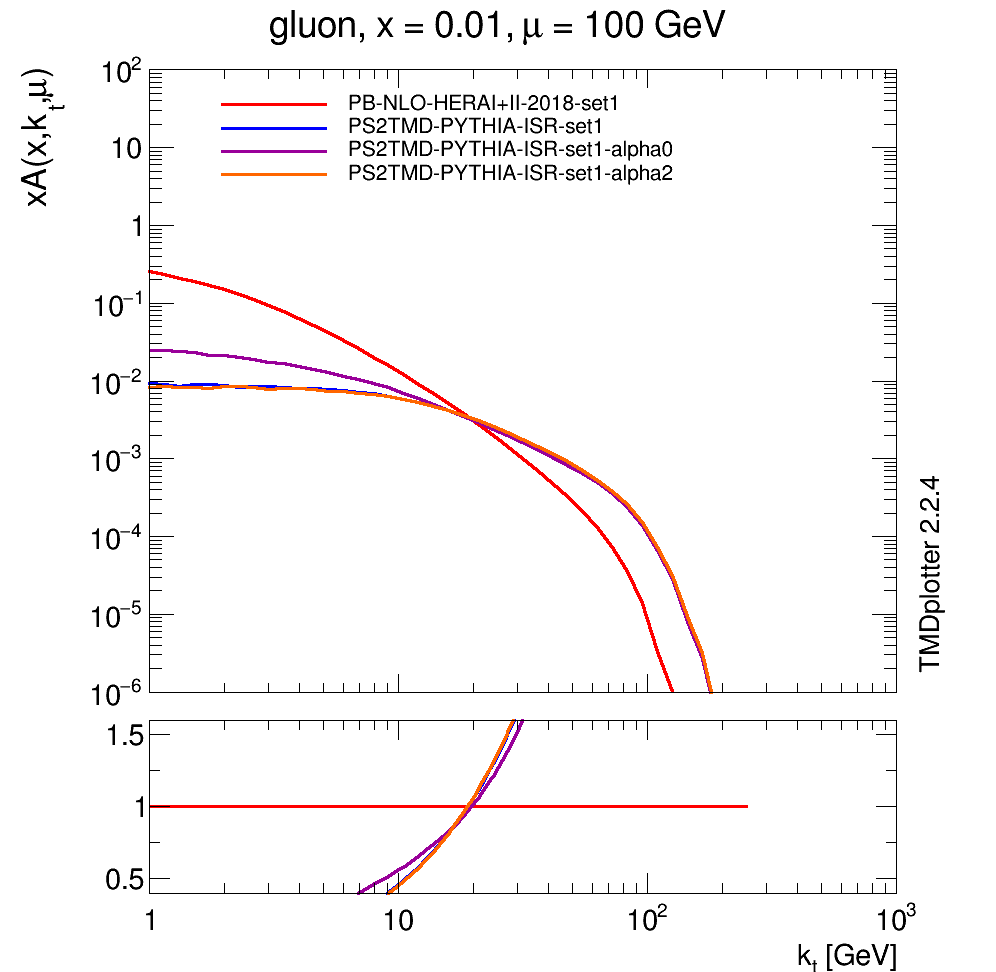}} 
    \subfigure[]{\includegraphics[width=0.42\textwidth]{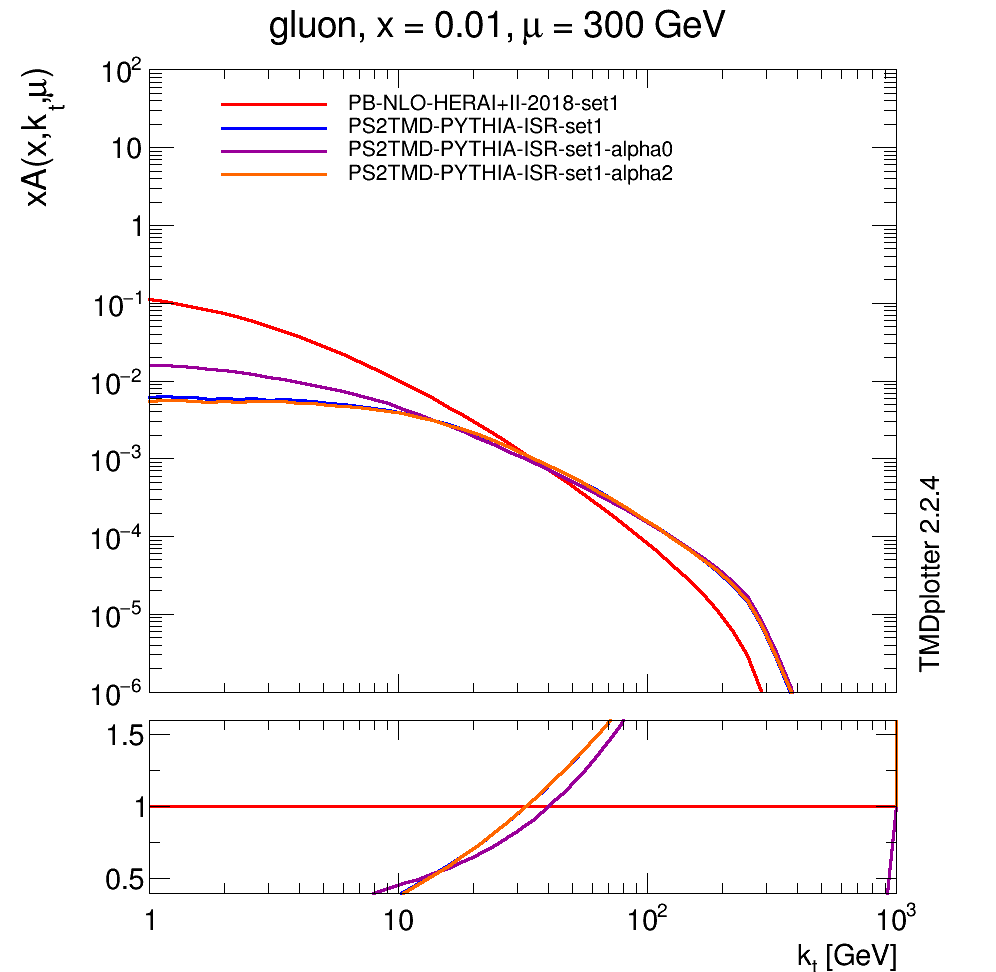}}
    \subfigure[]{\includegraphics[width=0.42\textwidth]{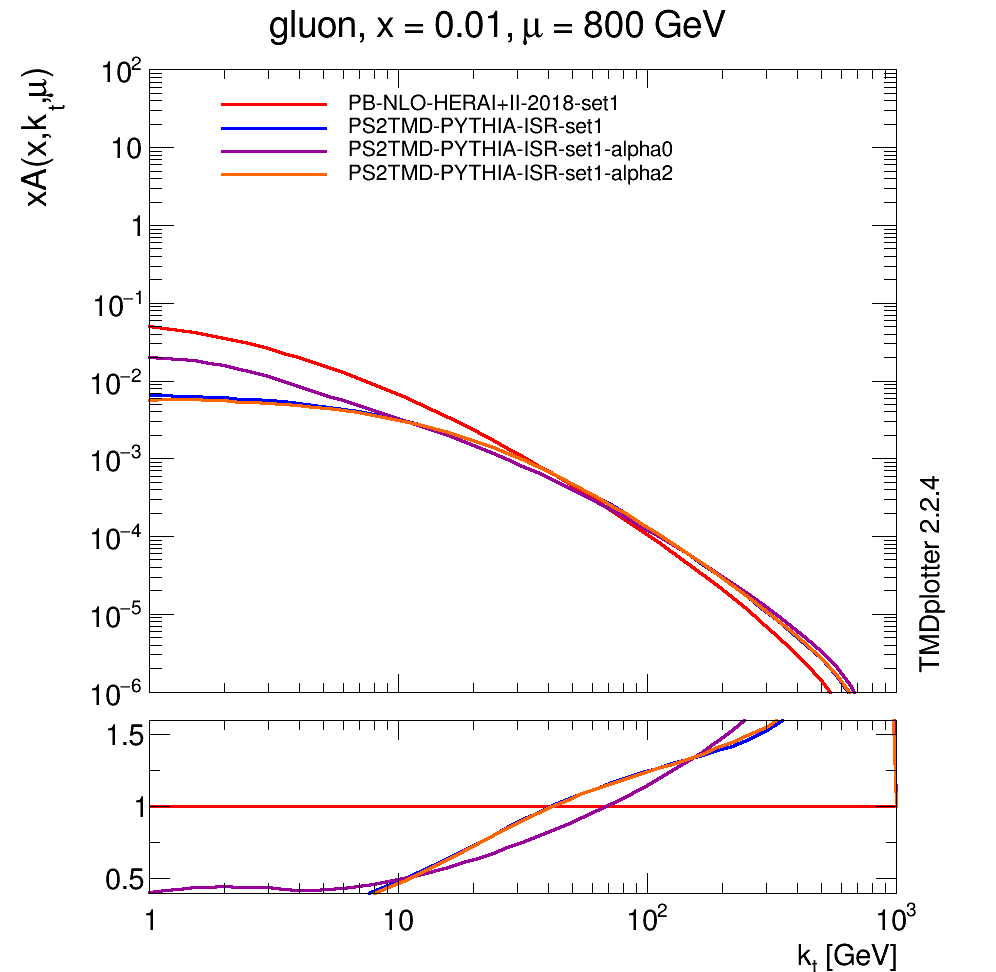}}
    \caption{Reconstructed TMDs for different \(\alpha_s\) orderings at (a) \(\mu\)= 10 GeV, (b) \(\mu\)= 100 GeV, (c) \(\mu\)= 300 GeV, (d) \(\mu\)= 800 GeV . }
    \label{fig:alpha}
\end{figure}

\paragraph{ISR + Rapidity Ordering OFF} Results are given in Figure \ref{fig:rap}. The rapidity ordering condition makes certain rapidity cuts and vetos within a defined range. When this requirement is switched off, those events that satisfies \(p_T\) ordering but not rapidity ordering are allowed. Hence, one observes the tail of ISR + Rapidity OFF TMD distribution being extended. We conclude that rapidity ordering configuration affects the tail of distribution. This effect is observed to reduce as \(\mu\) increases.
\paragraph{ISR + Different \(\alpha_s\) orders } The order of strong coupling \(\alpha_s\) is changed from default option while the QCD coupling is fixed to $\alpha_s(M_Z^2)=0.1365$. The value of this parameter indicates the order of \(\alpha_s\) being used: 0 for \(0^{th}\) order where \(\alpha_s\) is fixed ; 1 (default value) for Leading Order (LO) where \(\alpha_s\) takes the conventional value and changes with energy ; 2 for Next-to-Leading Order (NLO) where \(2^{nd}\) order loop expansion is included. 
As shown in Figure \ref{fig:alpha}, there is an obvious agreement using LO and NLO \(\alpha_s\) at all energy scales; 
a significant difference observed if we go from running  to fixed \(\alpha_s\).

\section{Longitudinal momentum shifts}
In the most current implementations of TMD formalisms in Monte Carlo simulation tools, cross sections are computed with on-shell initial partons (collinear PDFs), where transverse momentum is not included in the hard scattering calculation. After generation of the initial-state parton shower, the initial parton receives a transverse momentum. In order to satisfy energy-momentum conservation, the longitudinal momentum has to be adjusted. This may lead to notable kinematic shifts in the longitudinal momentum fraction \(x\).

Here, we try to investigate the effect of this energy-momentum conservation constraints on the energy fraction $x$.
 The longitudinal momentum of parton~2  can directly be read from the event record without further calculation ; can be reconstructed from mass of the produced system $x_2=\frac{m}{\sqrt{s}}e^{\pm y}$ ; from the lightcone momentum fraction $x_2=\frac{E+p_z}{(E+p_z)_{beam}}$ ; or from the Sudakov decomposition. We have 
\begin{eqnarray}
p_1=x_1 ~\tilde{p_1} +x_2~~\tilde{p_2}+k_{t1}\\ \nonumber
p_2=x_1 ~\tilde{p_1}+x_2~\tilde{p_2}+k_{t2}~,
\end{eqnarray}
where $p_1$ and $p_2$ are incoming partons four components and $~\tilde{p_1}$ and $~\tilde{p_2}$ are incoming proton's four momentum. The Sudakov decomposition longitudinal momentum fraction can be achieved via $x_2=\frac{p_2 . ~\tilde{p_1}}{~\tilde{p_2} . ~\tilde{p_1}}$. 

\begin{figure}[!h]
    \centering
    \subfigure[$x$ from PDF]{\includegraphics[width=0.42\textwidth]{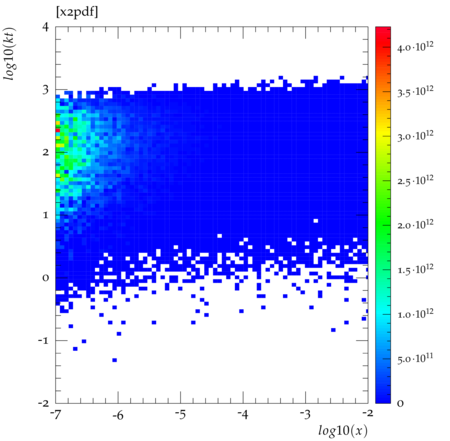}} 
    \subfigure[$x$ from mass ]{\includegraphics[width=0.42\textwidth]{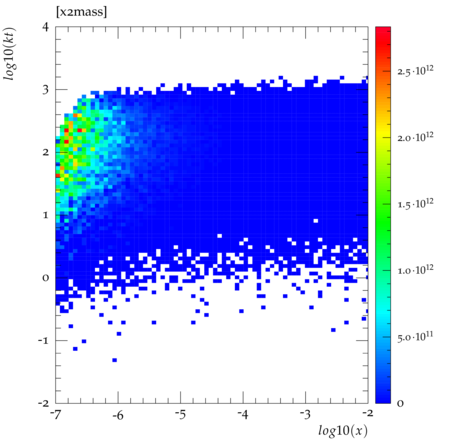}}
    \subfigure[$x$ from lightcone]{\includegraphics[width=0.42\textwidth]{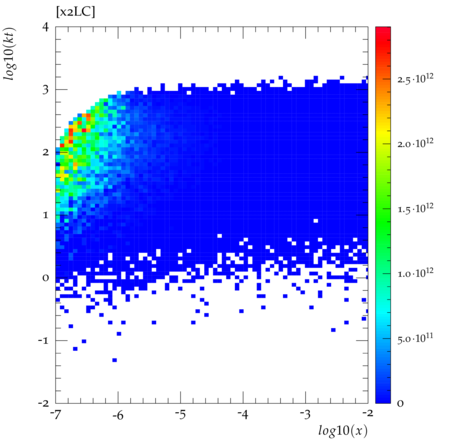}}
    \subfigure[$x$ from Sudakov]{\includegraphics[width=0.42\textwidth]{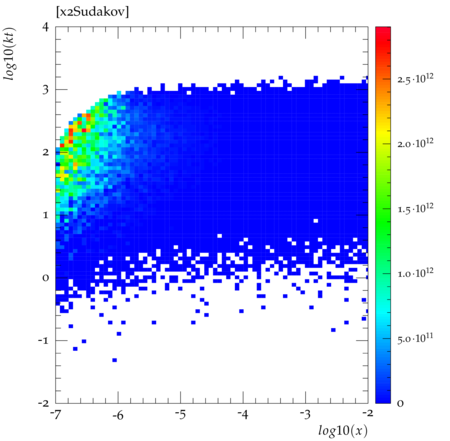}}    
    \caption{Reconstructed \(k_T\) versus \(x\) after initial-state parton showering for different \(x\) definitions: directly from HEPMC file (a), mass definition (b), Lightcone definition (c) and Sudakov decomposition (d).}
    \label{fig:diffX}
\end{figure}

In the collinear case these three definitions of $x_2$ are equivalent.
Turning on initial state parton shower, we reach slightly different $x_2$ values at very small \(x\) and large \(k_T\). These longitudinal momentum reshuffling can seriously affect the accuracy of the calculations in this specific phase space regions.
Figure~\ref{fig:diffX} shows the reconstructed \(k_T\) versus \(x\) after initial-state parton showering for different $x$ definitions.


%

\begin{figure}[!h]
    \centering
    \subfigure[]{\includegraphics[width=0.42\textwidth]{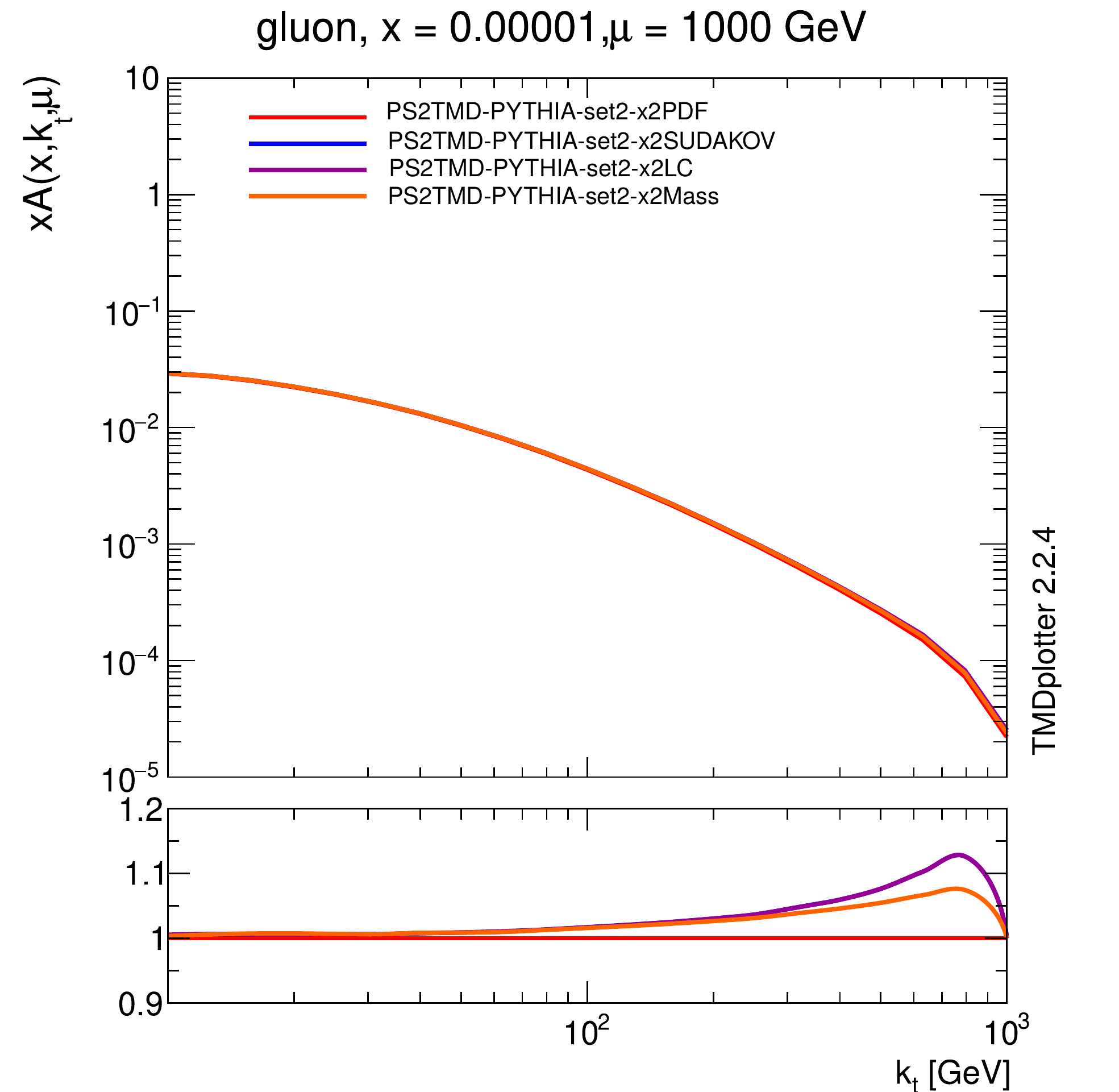}} 
    \subfigure[]{\includegraphics[width=0.42\textwidth]{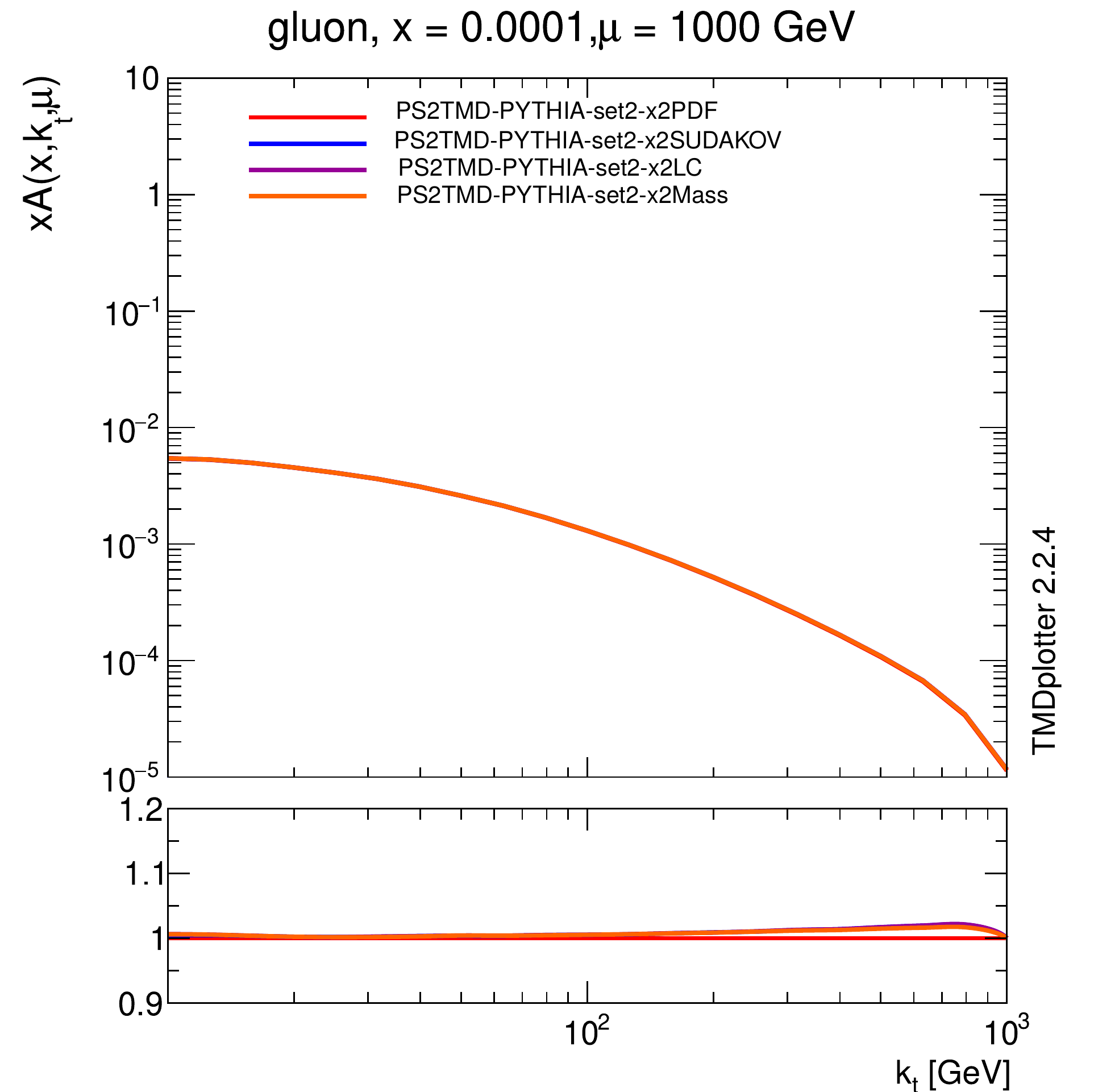}}
    \caption{gluon TMD distributions at $x=10^{-5},~10^{-4}$ and $\mu=10^3$ for different $x_2$ definitions.}
    \label{fig:diffXTMD}
\end{figure}

\begin{figure}[!h]
    \centering
    \subfigure[]{\includegraphics[width=0.42\textwidth]{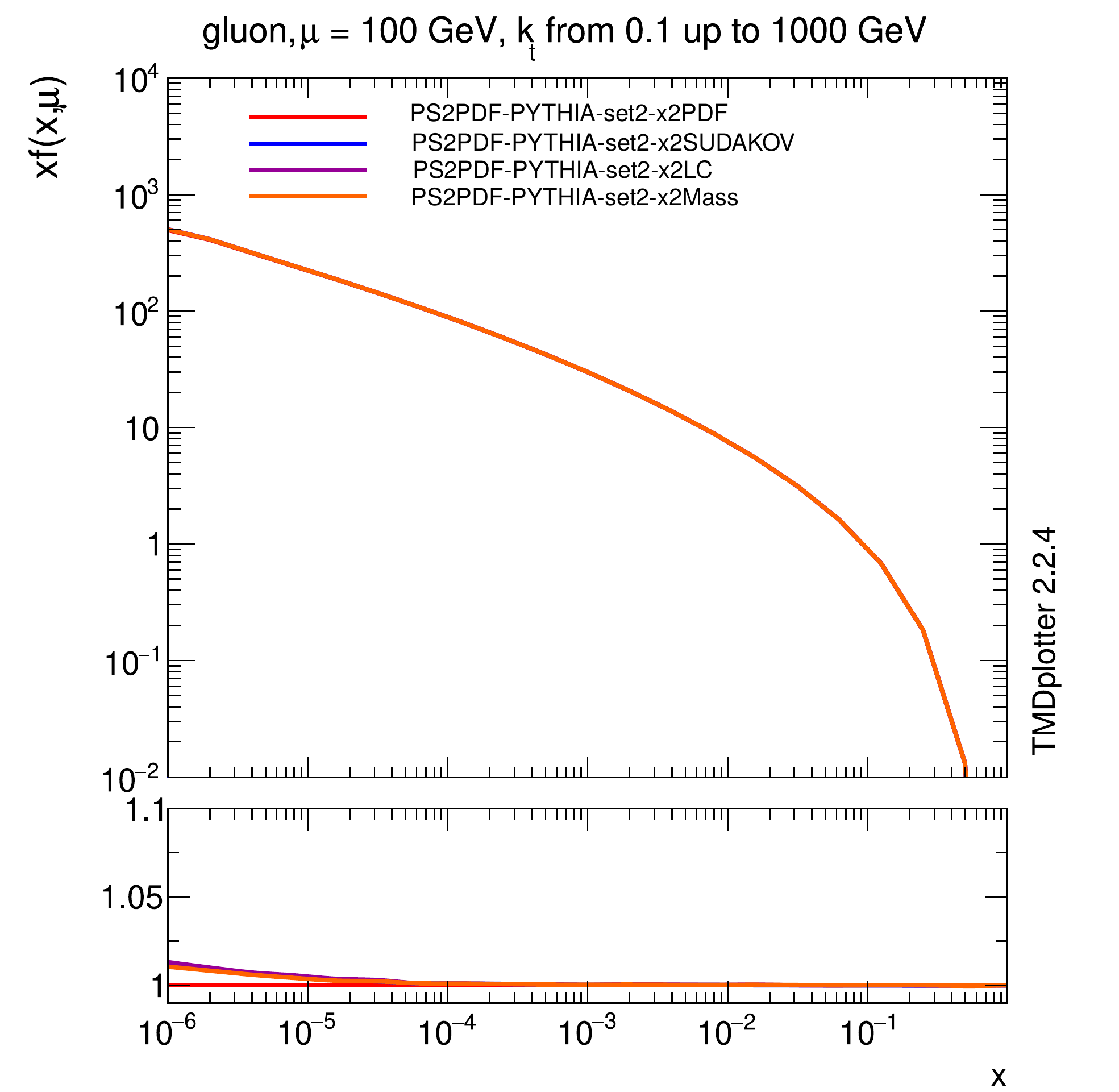}} 
    \subfigure[]{\includegraphics[width=0.42\textwidth]{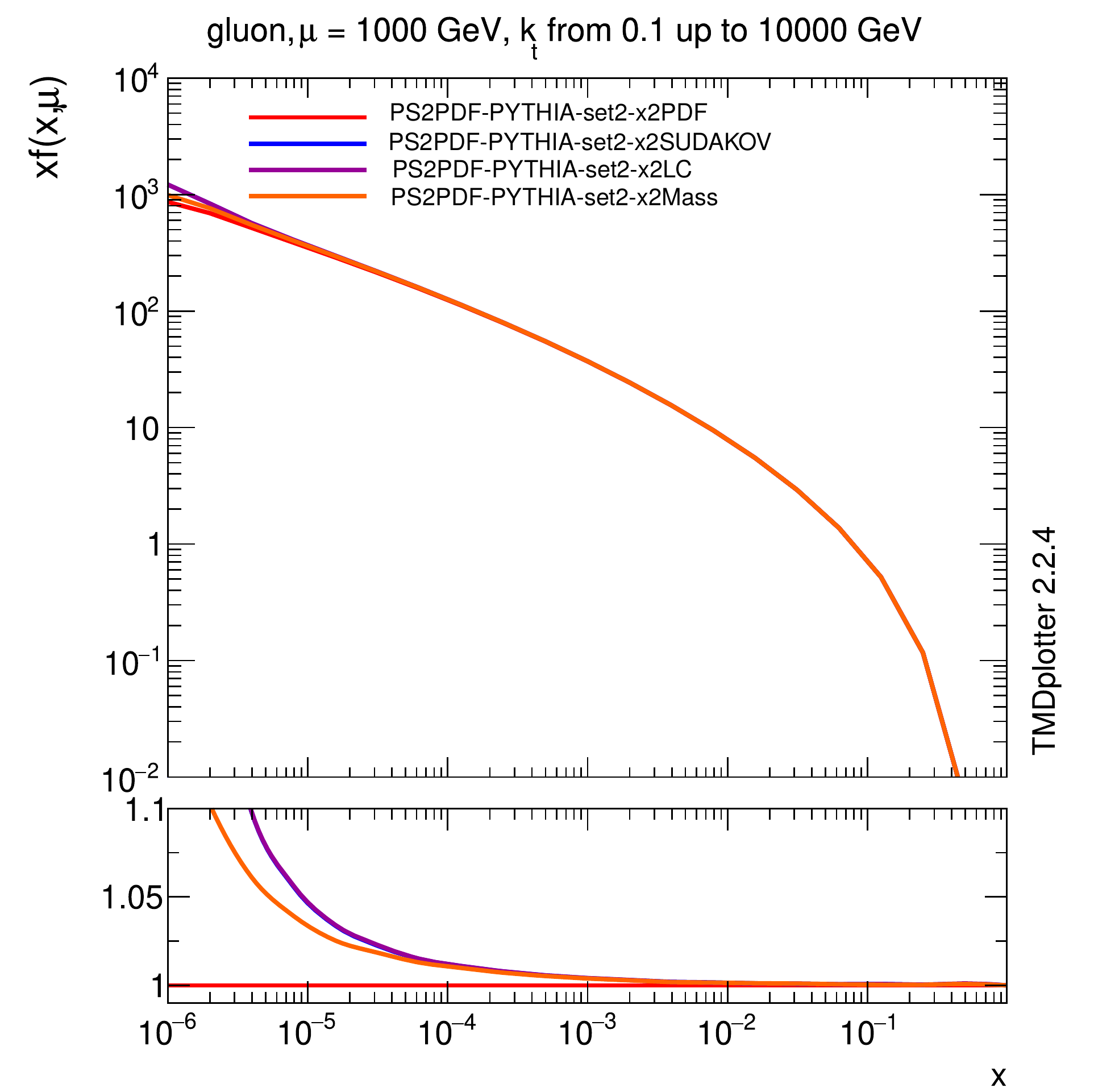}}
     \caption{gluon PDF distributions at  $\mu=10^2,~10^3$ for different $x_2$ definitions.}
    \label{fig:diffXITMD}
\end{figure}

We also checked the influence of different longitudinal momentum definitions on reconstructed TMD and collinear parton distributions in Figure~\ref{fig:diffXTMD} and Figure~\ref{fig:diffXITMD}, respectively.
These effects are visible at large scales and small $x$. 
This can be crucial for phenomenological studies while the theory is compared with the experimental measurements over a wide range in $x$. 
The kinematic reshuffling can be avoided by using $k_t$ from the beginning of the hard process. 


\section{Conclusion}
 
 
 We implemented a special method, PS2TMD, to reconstruct TMDs from the Monte Carlo parton showers. 
 The PS2TMD method was first validated by applying it to TMD sets with \cascade{} Monte Carlo generator. After different consistency checks, the method was applied to study how different \pythia{8} parton shower configurations affect the reconstructed TMDs.

We perform detailed studies on the effects of
using different definitions of Bjorken $x$ as one of the internal evolution variables. As a result, we found that depending on the definition of the longitudinal momentum, one gets different collinear and TMD PDFs in the specific phase space regions. 
The PS2TMD method offers a direct comparison of parton shower between different approaches without the complication of final state observables. The PS2TMD method has proven as a useful and accurate approach. 
 
 \clearpage
 \section{Acknowledgements}
STM thanks the Humboldt Foundation for the Georg Forster research fellowship. SS and YZ greatly acknowledge the opportunity provided by DESY to participate in the summer student program.

\providecommand{\etal}{et al.\xspace}
\providecommand{\href}[2]{#2}
\providecommand{\coll}{Coll.}
\catcode`\@=11
\def\@bibitem#1{%
\ifmc@bstsupport
  \mc@iftail{#1}%
    {;\newline\ignorespaces}%
    {\ifmc@first\else.\fi\orig@bibitem{#1}}
  \mc@firstfalse
\else
  \mc@iftail{#1}%
    {\ignorespaces}%
    {\orig@bibitem{#1}}%
\fi}%
\catcode`\@=12

%

\end{document}